\documentclass[conference]{IEEEtran}

\usepackage[bookmarks=false]{hyperref}
\usepackage{comment}
\usepackage{epsfig}
\usepackage{subfigure}
\usepackage{amsfonts}
\usepackage{amsmath}
\usepackage{amssymb}
\usepackage{graphicx}
\usepackage{epstopdf}
\usepackage{graphicx}
\usepackage{array}
\graphicspath{ {./images/} }
\usepackage{amsmath,amssymb,amsfonts}
\usepackage{algorithmic}
\usepackage{graphicx}
\usepackage{textcomp}
\usepackage{caption,setspace}
\usepackage{multirow}
\usepackage{listings}
\usepackage{url}
\usepackage{hyperref}
\makeatletter

\usepackage{listings, xcolor}

\definecolor{verylightgray}{rgb}{.97,.97,.97}

\lstdefinelanguage{Solidity}{
	keywords=[1]{anonymous, assembly, assert, balance, break, call, callcode, case, catch, class, constant, continue, constructor, contract, debugger, default, delegatecall, delete, do, else, emit, event, experimental, export, external, false, finally, for, function, gas, if, implements, import, in, indexed, instanceof, interface, internal, is, length, library, log0, log1, log2, log3, log4, memory, modifier, new, payable, pragma, private, protected, public, pure, push, require, return, returns, revert, selfdestruct, send, solidity, storage, struct, suicide, super, switch, then, this, throw, transfer, true, try, typeof, using, value, view, while, with, addmod, ecrecover, keccak256, mulmod, ripemd160, sha256, sha3}, 
	keywordstyle=[1]\color{blue}\bfseries,
	keywords=[2]{address, bool, byte, bytes, bytes1, bytes2, bytes3, bytes4, bytes5, bytes6, bytes7, bytes8, bytes9, bytes10, bytes11, bytes12, bytes13, bytes14, bytes15, bytes16, bytes17, bytes18, bytes19, bytes20, bytes21, bytes22, bytes23, bytes24, bytes25, bytes26, bytes27, bytes28, bytes29, bytes30, bytes31, bytes32, enum, int, int8, int16, int24, int32, int40, int48, int56, int64, int72, int80, int88, int96, int104, int112, int120, int128, int136, int144, int152, int160, int168, int176, int184, int192, int200, int208, int216, int224, int232, int240, int248, int256, mapping, string, uint, uint8, uint16, uint24, uint32, uint40, uint48, uint56, uint64, uint72, uint80, uint88, uint96, uint104, uint112, uint120, uint128, uint136, uint144, uint152, uint160, uint168, uint176, uint184, uint192, uint200, uint208, uint216, uint224, uint232, uint240, uint248, uint256, var, void, ether, finney, szabo, wei, days, hours, minutes, seconds, weeks, years},	
	keywordstyle=[2]\color{teal}\bfseries,
	keywords=[3]{block, blockhash, coinbase, difficulty, gaslimit, number, timestamp, msg, data, gas, sender, sig, value, now, tx, gasprice, origin},	
	keywordstyle=[3]\color{violet}\bfseries,
	identifierstyle=\color{black},
	sensitive=false,
	comment=[l]{//},
	morecomment=[s]{/*}{*/},
	commentstyle=\color{gray}\ttfamily,
	stringstyle=\color{red}\ttfamily,
	morestring=[b]',
	morestring=[b]"
}

\def\UrlAlphabet{%
      \do\a\do\b\do\c\do\d\do\e\do\f\do\g\do\h\do\i\do\j%
      \do\k\do\l\do\m\do\n\do\o\do\p\do\q\do\r\do\s\do\t%
      \do\u\do\v\do\w\do\x\do\y\do\z\do\A\do\B\do\C\do\D%
      \do\E\do\F\do\G\do\H\do\I\do\J\do\K\do\L\do\M\do\N%
      \do\O\do\P\do\Q\do\R\do\S\do\T\do\U\do\V\do\W\do\X%
      \do\Y\do\Z}
\def\UrlDigits{\do\1\do\2\do\3\do\4\do\5\do\6\do\7\do\8\do\9\do\0}
\g@addto@macro{\UrlBreaks}{\UrlOrds}
\g@addto@macro{\UrlBreaks}{\UrlAlphabet}
\g@addto@macro{\UrlBreaks}{\UrlDigits}
\makeatother
\ifCLASSOPTIONcompsoc
  \usepackage[nocompress]{cite}
\else
  \usepackage{cite}
\fi

\ifCLASSINFOpdf
\else
\fi

\begin{document}

\title{A Security Case Study for Blockchain Games}

\author{\IEEEauthorblockN{Tian Min}
\IEEEauthorblockA{\textit{School of Science and Engineering} \\
\textit{The Chinese University of Hong Kong, Shenzhen}\\
Shenzhen, China \\
tianmin@link.cuhk.edu.cn}
\and

\IEEEauthorblockN{Wei Cai}
\IEEEauthorblockA{\textit{School of Science and Engineering} \\
\textit{The Chinese University of Hong Kong, Shenzhen}\\
Shenzhen, China \\
caiwei@cuhk.edu.cn}
}

\maketitle

\begin{abstract}

Blockchain gaming is an emerging entertainment paradigm. However, blockchain games are still suffering from security issues, due to the immature blockchain technologies and its unsophisticated developers. In this work, we analyzed the blockchain game architecture and reveal the possible penetration methods of cracking. We scanned more than 600 commercial blockchain games to summarize a security overview from the perspective of the web server and smart contract, respectively. We also conducted three case studies for blockchain games to show detailed vulnerability detection.
\end{abstract}


\maketitle

\section{Introduction}\label{sec:introduction}

With the popularity of cryptocurrencies, e.g. BitCoin \cite{bitcoin}, the blockchain technology \cite{Nofer2017} is now recognized as the foundation of next-generation digital economy. However, the research community is looking forward to unleashing the full potential of blockchain for other businesses. To respond to the demand, Ethereum \cite{ethereum} brought smart contracts \cite{smartcontractinblockchain} to announce the start of Blockchain 2.0 era. A smart contract, written as immutable blockchain transactions, are transparent and auditable programs that can be automatically executed without any centralized control. With the support of the smart contract, decentralized applications (DApps) \cite{WeiCaiWEHFL2018} became possible.


Blockchain games become one of the most active DApp in the ecosystem. According to the recent survey work \cite{blockgamesurvey}, over 50\% of network traffic in Ethereum and EOS\footnote{https://eos.io/} platforms are from blockchain game players. The phenomenon can be explained from different perspectives. First, the non-fungible nature of blockchain data and the transparency of the smart contracts enable game developers to better prove the rule transparency, guarantee the ownership of the virtual asset, enable assets reusability, and encourage user-generated contents. Second, the blockchain game builds the whole ecosystem in the virtual world, which avoids a lot of realistic constraints commonly exist in other DApps, including the Internet of Things and source tracking. According to the above advantages, the blockchain game is considered an emerging trend. The industry has started its exploration on this topic by integrating traditional games with blockchain systems. Until January 2019 on Ethereum\footnote{https://www.stateoftheDApps.com/stats}, games contributed 1,113,516 transactions, which is 56.8\% of the total, and carried a transaction volume of 198,457 ETH, which is equal to 22 million dollars. In the same month, games have 42,210 active users, which is 24.3\% of all. From these statistic data, we can tell that blockchain games have already become an important component of DApps and have held a considerable market capitalization.

However, DApps are facing severe security issues. According to PeckShield 2018 annual report\footnote{https://www.huoxing24.com/newsdetail/20190128132648252411.html}, the economic losses caused by blockchain security in 2018 amounted to 2.238 billion dollars, which is 253\% of the 2017's. The blockchain security issues are mainly concentrated on the application layer and the contract layer, with 64 and 58 incidents respectively. The economic losses caused by these layers accounted for 98.87\% of all. To be more specific, on Ethereum, there were 54 security incidents. Most of them happened because of the issues from the exchange trading system, wallet website security, smart contracts vulnerabilities, and blockchain design defects. On EOS, There were 49 security incidents. Most of the attacks directly targeted on EOS smart contracts. In EOS smart contracts, the vulnerabilities can be easily reproduced in others.

In this work, we use the blockchain game, the most popular public blockchain application, to conduct the security study in DApps. The remainder of this paper is organized as follows: we briefly introduce the previous works on blockchain technology and security in Section~\ref{sec:blockchain}. Then, we illustrate the system architecture for the blockchain games and provide possible attack methods in Section~\ref{sec:architecture}. Next, an statistic of blockchain game security will be presented in Section~\ref{sec:security overview}. Afterward, we conduct security case studies in Section~\ref{sec:security case study} to do practical demonstrations. Section~\ref{sec:conclusion} concludes the article and envision the future of blockchain games.

\section{Related Work}\label{sec:blockchain}

\subsection{Blockchain and Games}

Blockchain is a decentralized database system with the characteristic of transparency, immutability and traceability. Different from the traditional database, public blockchain is a system maintained by the public, which can be accessed and verified by anyone around the world. To be more specific, it is a series of continuously growing blocks. Each block contains a cryptographic hash of the previous block, a timestamp, and its conveyed data\cite{Nofer2017}. Due to the existence of the cryptographic hash, blockchains are immutable. If a block on chain is modified, all the descendants of this block should be regenerated with new hash value. The blockchain data structure, together with the peer-to-peer (P2P) system and the proof-of-work (PoW) \cite{Hashcash2002} consensus model, forms the solid foundation for DApps.

According to this definition, we only discuss blockchain games with decentralized nature in this paper. Therefore, those blockchain games which follow the centralized and close source principle will not be investigated. For example,  the CryptoRabits\footnote{https://jiamitu.mi.com/home} developed by Xiaomi Inc. will be out of the scope of our work, since it was published without any public open source code or white paper. In addition, according to CryptoRabits' user agreement\footnote{https://jiamitu.mi.com/protocol}: 1. Users are not allowed to trade the currency in the game. 2. The operator has the right of making or adjusting the rules for the game. 3. If the user violates the agreement, the operator has the right to stop providing services to him immediately without his consent. These terms are completely contrary to the spirit of the blockchain. Players' ownership of virtual properties cannot be guaranteed.


\subsection{Smart Contract Vulnerability}

Initially proposed by Nick Szabo\cite{Szabo1997} in 1997, a smart contract is a protocol that can automatically verify and process the content of the contract. Thanks to the features of blockchain systems, the smart contract can ensure the code execution without the third-parties. Different blockchain platform may have different regulations on smart contracts programming. Nevertheless, all smart contracts have structures like object-oriented programs. On Ethereum, smart contracts are programs wrote in a JavaScript-like language called Solidity\footnote{https://solidity.readthedocs.io/en/develop/index.html}. Each contract is like a class, which contains variables and methods. Contracts can also invoke each other to implement complicated tasks. Following is a simple example of a Solidity smart contract.

\begin{lstlisting}[language=Solidity, numbers=none]
pragma solidity >=0.4.0 <0.7.0;

contract SimpleStorage {

    uint storedData;

    function set(uint x) public {
        storedData = x;
    }

    function get() public view returns (uint) {
        return storedData;
    }
}
\end{lstlisting}

However, a smart contract based program is susceptible. A study \cite{Nikolic2018} pointed out that there are 34,200 contracts marked as vulnerable in a million samples. Some of the most essential reasons for the vulnerabilities are platform and the contract programming language's design defects. Take Ethereum and Solidity as an example. Most of the currently known vulnerabilities about the smart contracts are related to the fallback function, which is an unnamed function triggered when an external caller is sent ETH, the Ethereum token, to the contract, or calls a function that does not defined. When \textit{fallback()} includes an external function or has potential vulnerabilities, the attackers could hijack the invoked contract, and force it to execute.


\subsection{Smart Contract Audition Tools}
Code auditing is not a new concept. It is an integral part of the defensive programming paradigm, which attempts to reduce errors before the software is released. However, Solidity is a high-level programming language with many potentially vulnerable functions. Due to the unchangeability of the blockchain, updating patches after deployment becomes especially troublesome. Hence, the smart contract audition becomes of paramount importance. As system developers and operators are gradually aware of the importance of blockchain security, more and more auditing tools have emerged. Different tools may have different advantages including automation degree, accuracy and efficiency. Audition tools detect the vulnerabilities in three main ways: 1) Code Feature Matching: Auditor collects and extracts malicious code's feature, and do semantic matching on other source code. 2) Formal Verification: Formal Verification is mathematical access to prove a system's completeness. Auditor specifies every possible input and exhaustive every situation that might happen. 3) Symbolic Execution: Auditor generate a control flow graph by contract's logic units (like determining statements). From this logic graph, The auditor can traverse all codes paths to reveal how the variables passing through the program in order to detect logical design flaws.



\section{Blockchain Game Architecture}\label{sec:architecture}

\begin{figure}[htp]
\centering
\includegraphics[width=9cm]{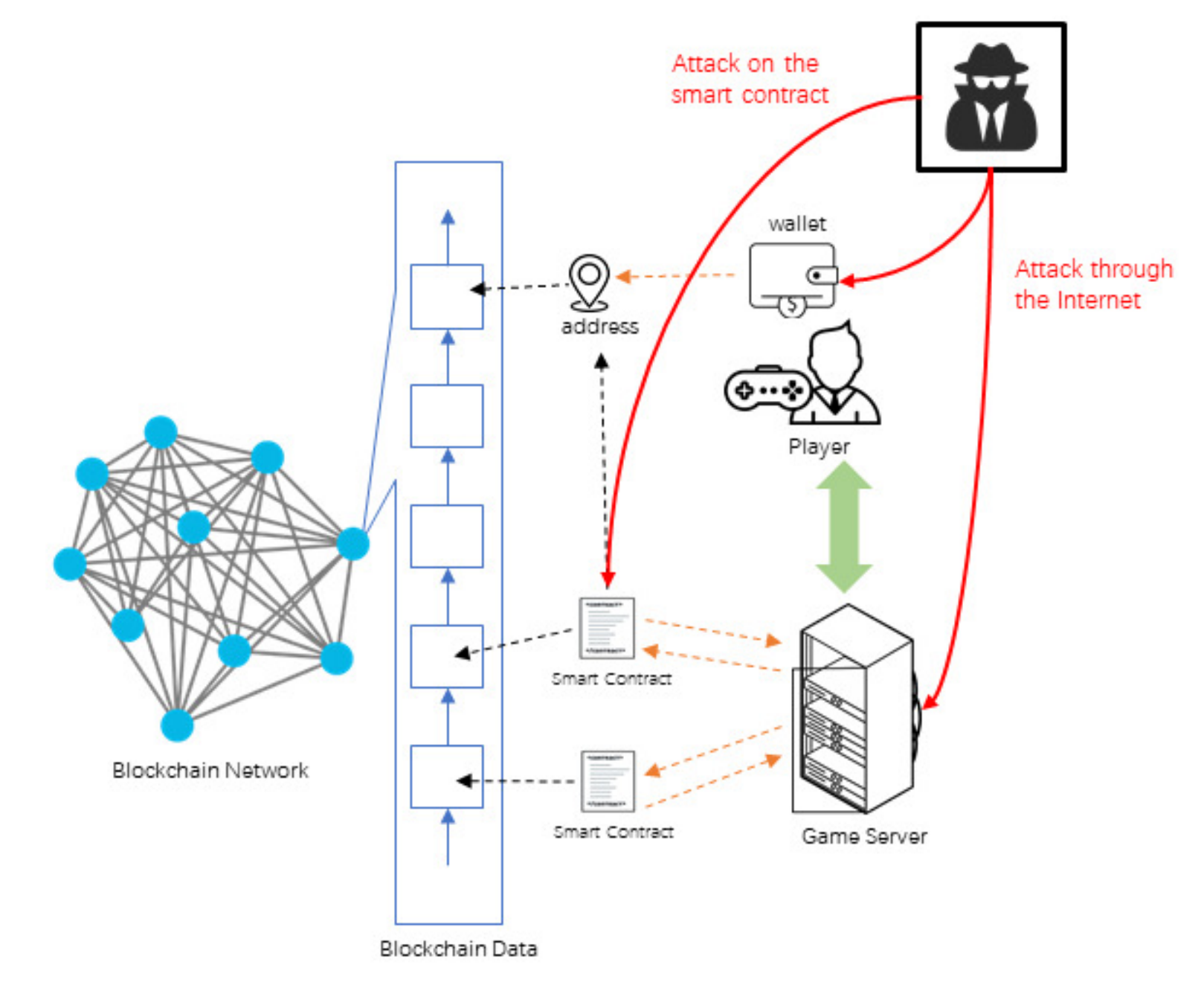}
\caption{Architecture for Blockchain Games}\label{fig:architecture}
\end{figure}

Fig. \ref{fig:architecture} illustrates the architecture of conventional blockchain game. Different from the traditional games, blockchain game players need to register an address in the corresponding blockchain platform before starting their gaming sessions. This blockchain address, accessed by a wallet program, will serve as a unique identity and a destination of the virtual assets for its corresponding player. On the other hand, the game server should offload some core functions, e.g. the ones which manipulate the players' virtual assets or critical game rules, to the blockchain as smart contracts in order to keep them transparent and immutable.

The server plays an important role in this architecture. In addition to providing game service, it acts as a cache and indexing engine for smart contracts. Although the ultimate source of information is from the blockchain, clients rely on the server's searching and verifying capabilities of the data returned from the blockchain. Moreover, writing data to PoW blockchains, e.g. Ethereum, is expensive. The server still needs to store most of the data and only store a hash on the chain for verification. The server of a blockchain game interact with the Ethereum blockchain via web3.js, which is a collection of libraries which allow developers to interact with a local or remote Ethereum node, using an HTTP, WebSocket or IPC connection\footnote{https://github.com/ethereum/wiki/wiki/JavaScript-API}. According to the architecture, there are mainly two methodologies of attacking a blockchain game:

\subsection{Web}
The focal point of penetrating a blockchain application is accessing the digital assets. Hence, the wallet becomes of great importance. The private key to the wallet shall be the optimal target of the attackers. Once the attacker obtains the private key, he could easily transfer assets away if there is no a two-Step verification. The secondary target shall be the game server since it is where the susceptible information may be stored. These information could help the attacker with further penetration. The ideology of penetrating a blockchain wallet or a DApp server may not has a great difference from traditional cyber attack.

\subsection{Smart Contract}
 Since all contracts are open source, the attackers can identify the vulnerable spots directly by analyzing the source code. Although most smart contracts were compiled into bytecodes before deployment, there are various tools that help reverse engineering. Smart contract vulnerabilities may exist in many different layers, including Solidity language, execution logic, and Ethereum Virtual Machine (EVM) design. in TABLE: \ref{smvulner}, Nicola Atzei\cite{SurveySmartContractAttack} summarized a taxonomy of smart contract vulnerability. It shows that the vulnerable spots can be found through the entire work-flow of smart contract execution.

\begin{table}[htp]
\centering
\begin{tabular}{|l|l|}
\hline
\textbf{Level} & \textbf{Cause of vulnerability} \\ \hline
\multirow{6}{*}{Solidity} & Call to the unknown \\ \cline{2-2}
 & Gasless send \\ \cline{2-2}
 & Exception disorders \\ \cline{2-2}
 & Type casts \\ \cline{2-2}
 & Re-entrancy \\ \cline{2-2}
 & Keeping secrets \\ \hline
\multirow{3}{*}{EVM} & Immutable bugs \\ \cline{2-2}
 & Ether lost in transfer \\ \cline{2-2}
 & Stack size limit \\ \hline
\multirow{3}{*}{Blockchain} & Unpredictable state \\ \cline{2-2}  & Generating randomness \\ \cline{2-2}
 & Time constraints \\ \hline
\end{tabular}
\caption{Smart Contract Vulnerability}
\label{smvulner}
\end{table}

\section{Statistics of Security Risks}\label{sec:security overview}
We selected 610 games listed on the State-of-the-DApps\footnote{https://www.stateoftheDApps.com/} and collected URLs and smart contract codes for analysis. Nikto2\footnote{https://cirt.net/Nikto2}, a web scanner, was employed to detect the vulnerabilities on the server and the web application, and Mythril\footnote{https://github.com/ConsenSys/mythril-classic} was used to detect vulnerabilities in their smart contracts.

\subsection{Web Overview}

\begin{figure}[htp]
\centering
\includegraphics[width=8.8cm]{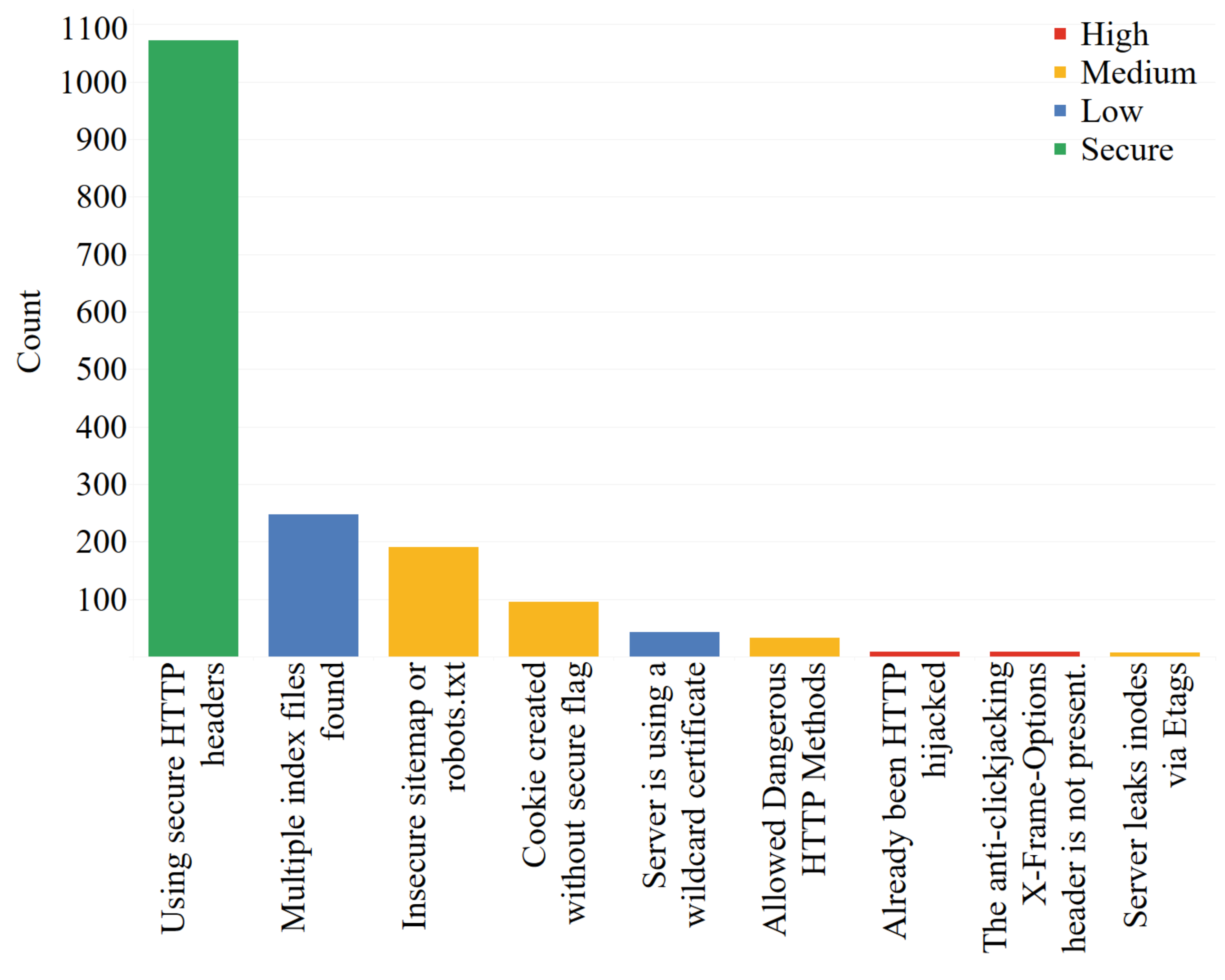}
\caption{Web App Issue Count and Severity}\label{fig:web_overview}
\end{figure}

Nikto2 is not an aggressive scanning tool. It mainly detects the misconfiguration on the web servers and the protocols. In 610 sites, more than 1,700 URLs, mainly 8 kinds of issues were detected. The result in Fig. \ref{fig:web_overview} shows that 62.91\% of the samples are using secure HTTP headers like "Strict-Transport-Security" or "X-XSS-Protection" to prevent being attacked. 16.96\% of them are under low-level risk, having multiple index files or using a wildcard certificate, which means if one server or sub-domain is compromised, others may also be exposed to the danger. 19.41\% of the samples are at medium level risk, they allow some risky HTTP methods, create cookies without a secure flag, or leaving sitemaps may have potential vulnerabilities. Rest 0.72\% of them have already been hijacked or haven't opened anti-clickjacking X-Frame-Options.

From the statistics, we can draw a conclusion that there has been systematic security development frameworks or toolkit for the developers when they started to build their own web server, thanks to the long-term development of web security. Followings we briefly introduce some common potential attacks which could be caused by these vulnerabilities:

\textbf{Cookie Replay} - caused by Cookie created without secure flag: The secure flag is an option that can be set by the application server when sending a new cookie to the user within an HTTP Response. This flag can prevent cookies from being observed by unauthorized parties due to the transmission of a the cookie in clear text. This vulnerability could allow the attacker impersonate the user as long as the cookie remains valid.

\textbf{Injection} - caused by Allow Dangerous HTTP Methods: Almost any data source can be an injection carrier, including environment variables, parameters, and external or internal web services. Blockchain games usually have frequent interaction with players, including lots of input box and complicated URL routes.

\textbf{XSS} - caused by X-Frame-Options not present: Cross-site scripting(XSS) is a kind of common vulnerabilities that happen on the web applications. Although most sites know to protect themselves with special filters, XSS attacks can be considered dangerous because they usually act as a springboard towards users' private keys. Secure headers like ``X-Frame-Options" must be included to eliminate html script ``<iframe>", which could lead to a clickjacking and fool players to input their passwords.

\textbf{Broken Authentication}: Apart from the scanning results above, directly cracking the authentication key is an important kinds of attacks. Attackers can do social engineering and use brute force to crack a wallet. Apart from weak passwords, poor session management also causes broken authentications. Especially for those sites exposing session ID in the URL, or creating token without encryption.

\subsection{Smart Contract Overview}

\begin{figure}[htp]
\centering
\includegraphics[width=8.8cm]{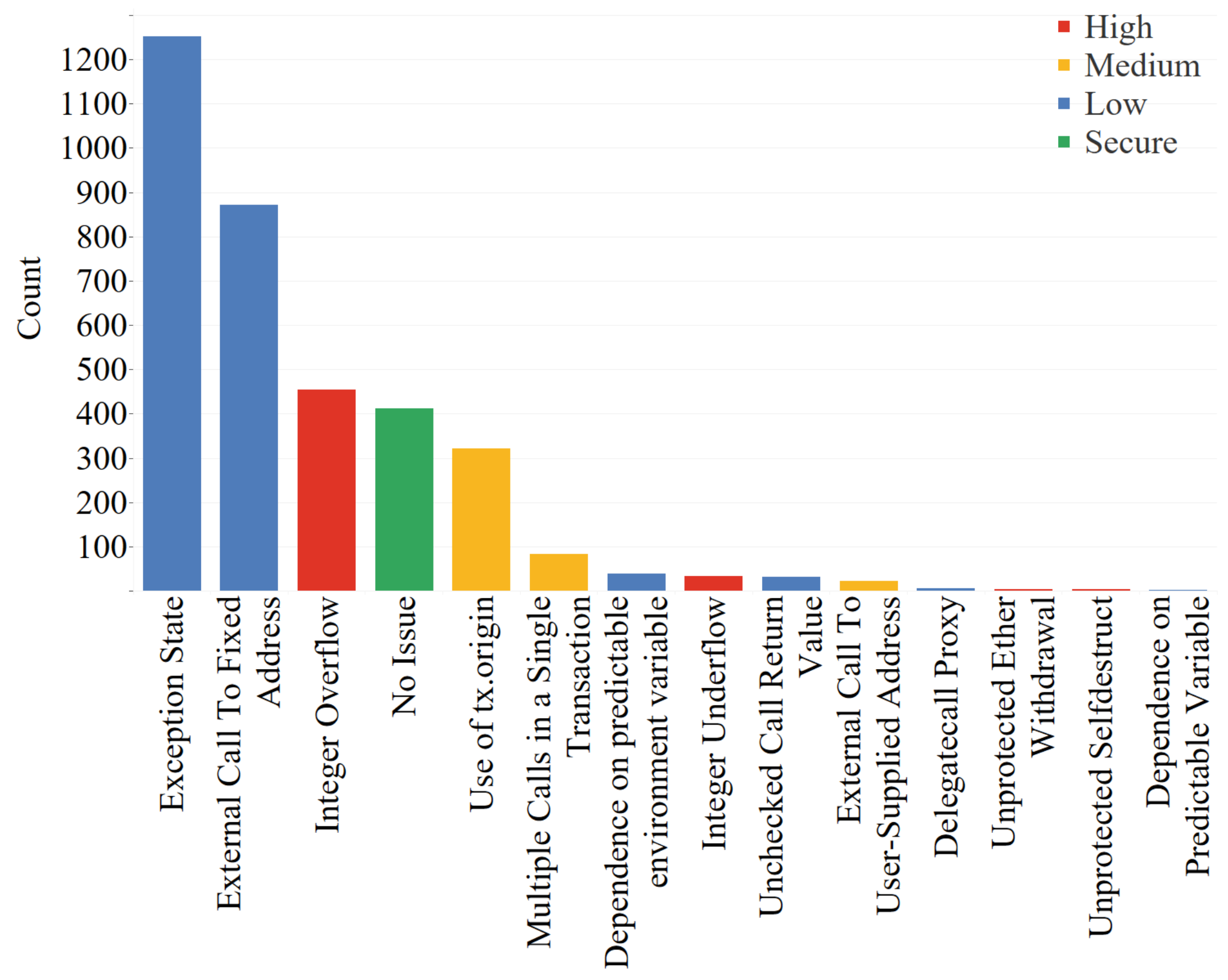}
\caption{Smart Contract Issue Count and Severity}\label{fig:sm_overview}
\end{figure}


In 1,311 smart contracts, 12 kinds of vulnerabilities are detected as the result are shown in Fig. \ref{fig:sm_overview}. Only 11.63\% contracts are bug-free. 14.04\% contracts have high-risk vulnerabilities like overflow and underflow, unprotected Ether withdrawal or unprotected self-destruct. 12.08\% of them are medium risks like usage of \textit{tx.origin} or applying multiple calls in a single transaction. Rest 62.25\% of the contracts are in low-level risk, having flaws like exception state or allowing external calls.

Among all the results, "Exception State" possesses the highest proportion of 35.43\%, which means that a large proportion of smart contract developers didn't handle the exceptions. The most common high severity vulnerability is "Integer Overflow", which take a proportion of 12.87\%.

We list some most common attacks that could be caused by the scanning results blow. You can check more up-to-date known attack on the Smart Contract Weakness Classification Registry\footnote{https://smartcontractsecurity.github.io/SWC-registry}:

\textbf{Overflow/Underflow} - caused by Integer Overflow/Underflow: Integer overflow and underflow are one of the most common vulnerabilities in the smart contract. If a UNIT256 reaches the maximum unit value($2^{256}$), it will circle back to zero when it's added with another 1. Attackers can exploit this vulnerability by repeatedly invoking the function that increases value. Game data like character's attack, defense or health point could be modified frequently, blockchain game developer may put special attention on value control.

\textbf{Tx Origin Attack} - caused by Use of \textit{tx.origin}: \textit{tx.origin} is a global variable that return the address which initially invoked the smart contract. Some developer use this variable to do the authentication. For example, the attacker wants to pass an authentication in a target contract. First, he might find a way to trick the victim into transferring to the malicious middle contract. Then, the middle contract will call the target contract in its \textit{fallback()}. So, the tx.origin address in the target contract should be the victim's address and authenticated.

\textbf{Predictable Variable} - caused by Dependence on predictable environment variable: Some smart contracts write functions that depend on variables like timestamp or block's height(distance from the genesis block), which are predictable. For example, when a blockchain casino generate dice points using the timestamp. The attacker can easily crack these functions and win the game.

\textbf{Denial of Service (DoS)}: DoS is a kind of common vulnerability which closely related to the fallback function and revert mechanism. The attacker can create a dead loop by logic vulnerabilities in the contracts. Take a bidding system as an example, if the system can only set a new bid leader after refunding to the previous one, the attacker can write a function that reverts any transaction in \textit{fallback()}, in order to keep being the bid leader.

\textbf{Re-entrancy}: Re-entrancy is a serious issue related to calling malicious external contracts, which may take over the control flow. This kind of vulnerabilities varies in different forms: every external call can be potentially dangerous. Nowadays, blockchain games are getting more and more complex, calling external smart contracts is unavoidable. If a developer must do an external call, he should cautiously verify the contract author, and try to arrange the call after the execution of the internal code.

Apart from making malicious calls to those vulnerable contracts, attackers can also take advantages of Ethereum's logical design flaws. Most of these flaws are dilemmas. For example, Gas is of great necessity to the PoW consensus model. However, attackers can manipulate the transaction by offering an expensive Gas fee. For example, Fomo3D is a gambling game with a 24-hour countdown. 30 seconds will be added every time when a token is sold. When the countdown touches 0, the last token buyer wins the jackpot. Due to the Ethereum's Gas mechanism, the attacker can do several transactions with expensive gas to jam the mining system, so thqt he/she could keep getting the top priority of the blockchain packing queue. As a result, other buyers' transaction cannot be successfully verified and written into blocks.

\section{Case Studies}\label{sec:security case study}

In this section, we conducted case studies to demonstrate the security issues in current commercial blockchain games. The first three cases are historical accomplished attacks, including EOSFomo 3D\footnote{https://eosfo.io}, Pandemica\footnote{https://pandemica.online/} and EOSlots\footnote{https://www.eoslots.com/}. We analyze their vulnerabilities and methods the attackers used. The rest of three cases are scanning result analysis, including
Cryptokitties\footnote{https://www.cryptokitties.co/}, 0xUniverse\footnote{https://0xuniverse.com/} and Mythereum\footnote{https://www.mythereum.io/}. We will showcase their high and medium level vulnerabilities in terms of web application aspect and smart contracts aspect.

\subsection{EOSFomo 3D}

\begin{figure}[htp]
\centering
\includegraphics[width=8.85cm]{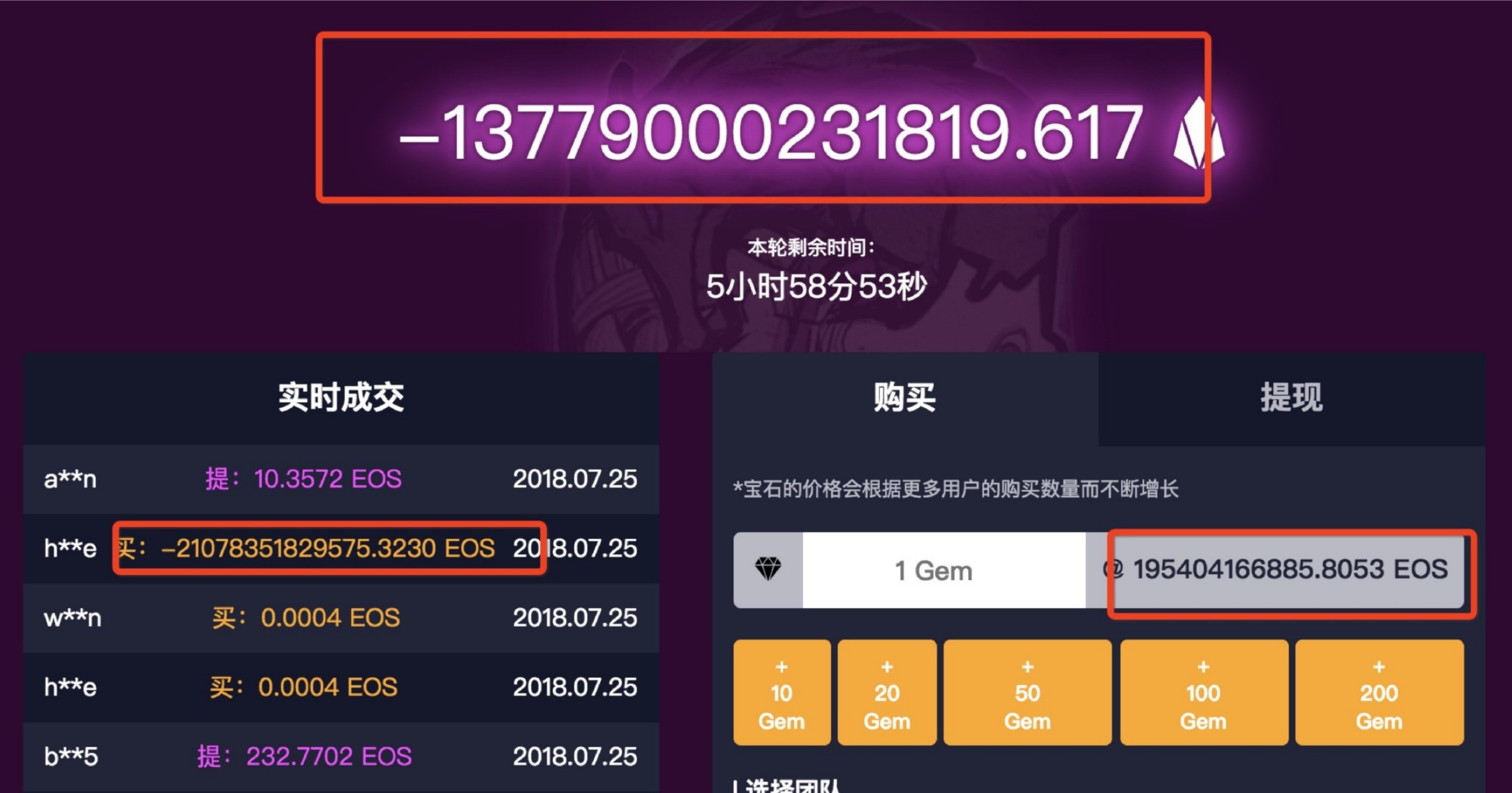}
\caption{EOSFomo 3D's Hompage After the Attack}
\label{EOSFomo}
\end{figure}

EOSFomo 3D is a Fomo3D\footnote{https://exitscam.me}-like game based on EOS platform. The players purchase keys on different teams and the last one receive rewards from jackpot. In July 2018, EOSFomo was attacked through an overflow vulnerability. As shown in Fig. \ref{EOSFomo}, the bonus displayed on the website became negative after the attack. In this incident, 60,686 EOS were stolen from ordinary users.

Vulnerability - \textbf{Overflow/Underflow}:
Because the game has already been shut down. We are unable to analyze its source code. However, we can infer that the developers did not verify the results or use a secure library. As shown in Fig. \ref{EOSFomo}, the overflow is triggered by a player, which means the rights management system has design flaws. The player can exploit an overflows by repeatedly calling a public function in the contract.

\subsection{Pandemica}

\begin{figure}[htp]
\centering
\includegraphics[width=8.85cm]{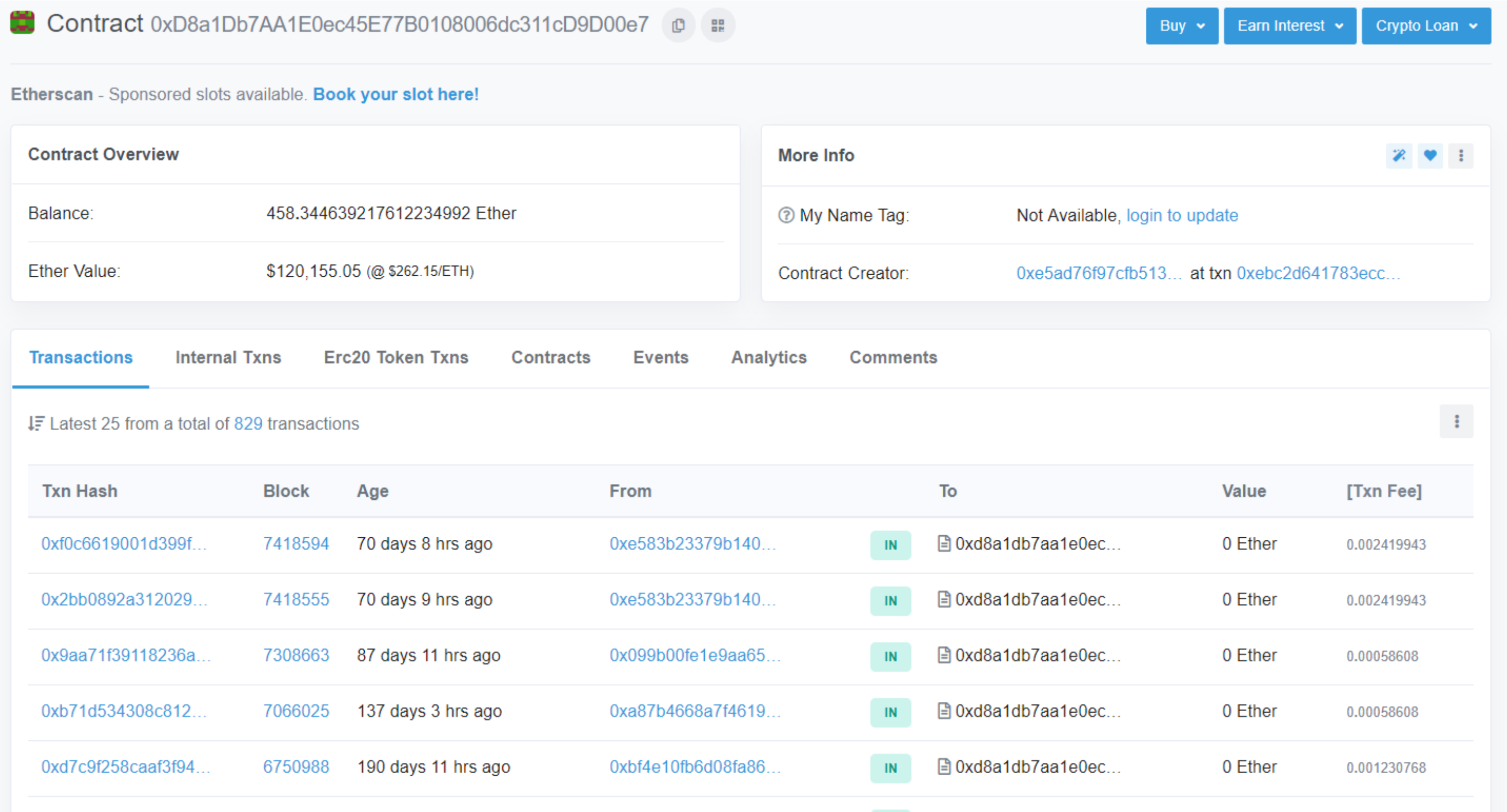}
\caption{Pandemica Contract on Etherscan}
\label{Pandemica}
\end{figure}

Pandemica is a Ethereum-based game following a simple Ponzi Scheme: players transfer ETH to the contract, and the owner randomly return 3\% of the collected fund to the players at 6:00 p.m. everyday. In August 2018, ETH worth 120 thousand USD was frozen in this contract\footnote{https://etherscan.io/address/0xd8a1db7aa1e0ec45e77b0108006dc311cd9d00e7} shown in Fig. \ref{Pandemica}.

Vulnerability - \textbf{Gas Overflow}: The contract developer used a loop to implement the paying method to the users:

\begin{lstlisting}[language=Solidity, numbers=none]
function Count() onlyowner {
        while (counter>0) {
            Tx[counter].txuser.send((Tx[counter].txvalue/100)*3);
            counter-=1;
        }
    }
\end{lstlisting}

The number of loops is determined by the number of participants. However, the amount of Gas that can be consumed in each block has a upper limit of 8,000,000 Gas.The \textit{counter} variable will grow as the increase of players. When the value of \textit{counter} reaches a certain threshold, Gas fee executing the \textit{Count()} function will exceed 8,000,000. This fund can only unfreeze when the Ethereum raise the upper bound of the Gas fee in the future.

\subsection{EOSlots}

\begin{figure}[htp]
\centering
\includegraphics[width=8.8cm]{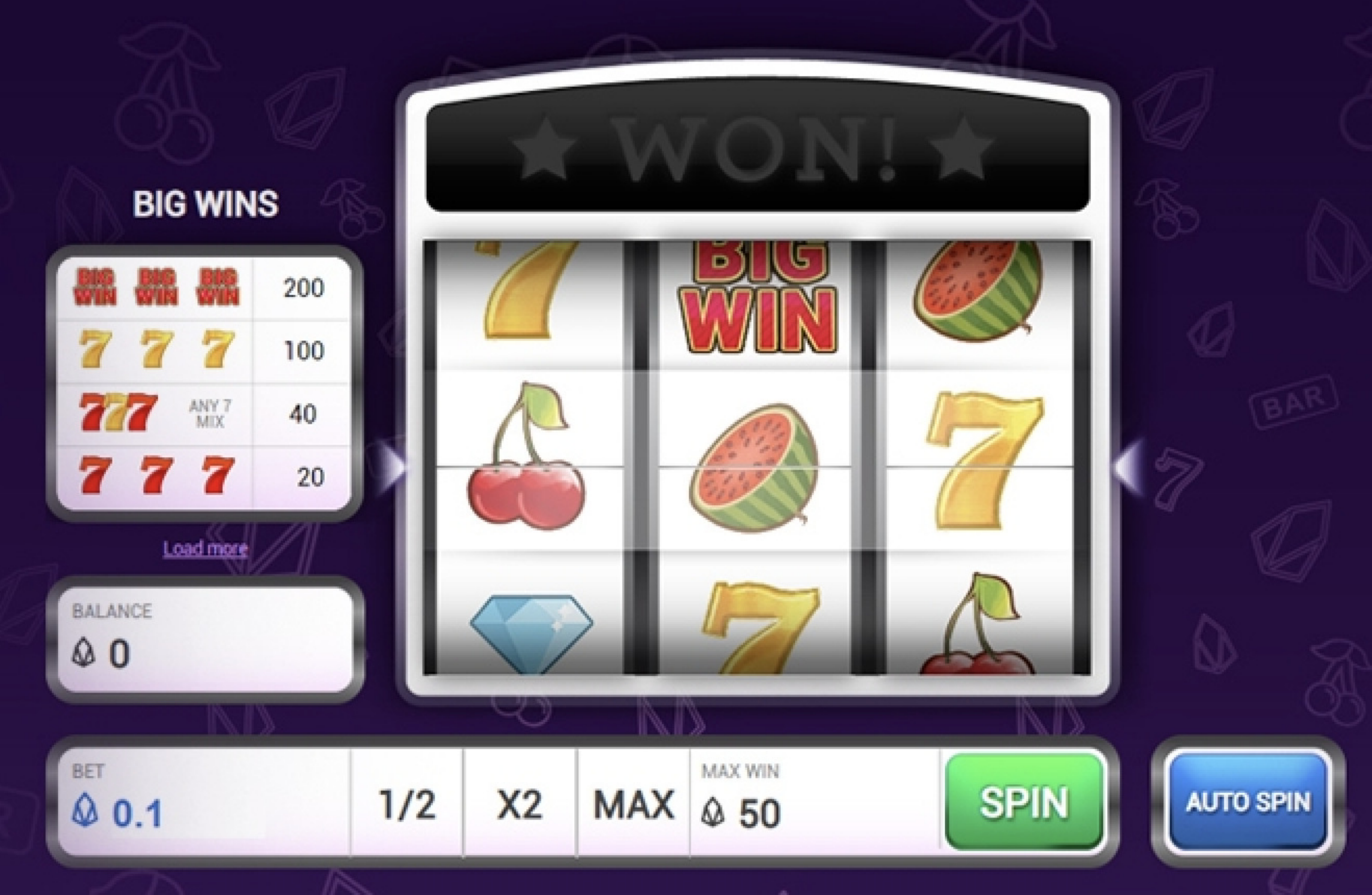}
\caption{Screenshot of EOSlots}
\label{EOSlotsScreenshot}
\end{figure}

EOSlots is a slot machine game on EOS platform as illustrated in Fig. \ref{EOSlotsScreenshot}. The developers claim it to be a fee-less and trust-less game where players can place bets in EOS at zero cost and have absolute certainty the game is fair, since the player’s funds go directly into a smart contract without the need for a middleman.

\begin{figure}[htp]
\centering
\includegraphics[width=9.1cm]{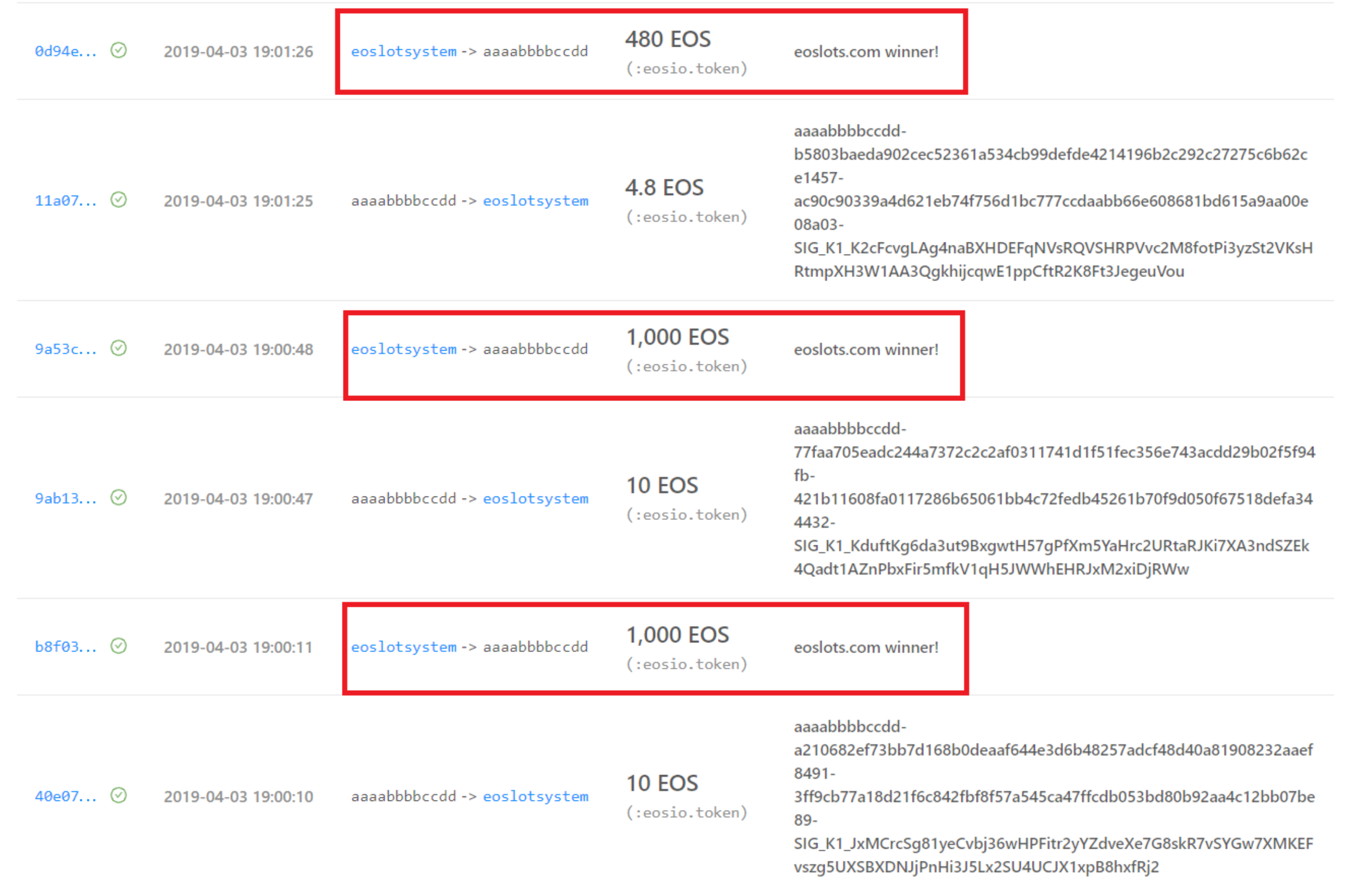}
\caption{Attacker aaaabbbbccdd Kept Winning the Game}
\label{EOSlots}
\end{figure}

As shown in Fig. \ref{EOSlots}, on April 3rd 2019, the attacker cracked the pseudo random number of EOSlots\footnote{https://eoslots.com/} and kept winning the game illegally. The attacker got ten times of the value he bet.

Vulnerability - \textbf{Predictable Variable}: Currently, Ethereum and EOS officials didn't provide a standardized random number interface, which causes a negative impact on blockchain games, especially the lottery games. In order to implement a random number generator, developers have to write their own functions, which often use the current block information as generator parameters. However, the attackers can generate the exact same value from the same parameter. Attackers can deploy a testing contract to keep generating random numbers and join the game after they got the numbers to look like the correct results.

\subsection{Cryptokitties}
CryptoKitties is a blockchain game developed by Axiom Zen that allows players to purchase, collect, breed and sell various types of virtual cats. It is one of the earliest and most successful blockchain games on Ethereum.

\begin{figure}[htp]
\centering
\includegraphics[width=9.1cm]{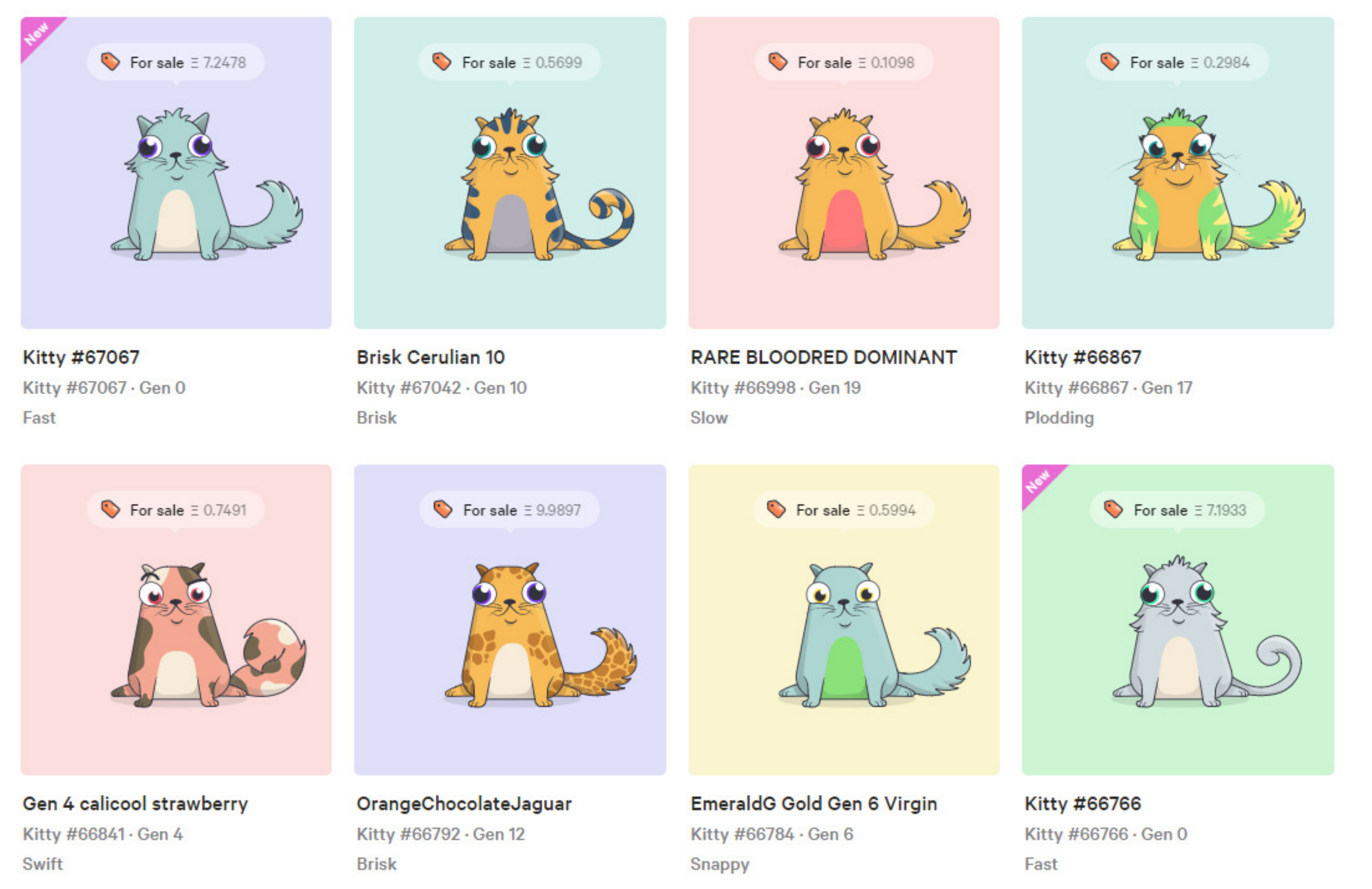}
\caption{Screenshot of Cryptikitties}\label{fig:cryptokitties}
\end{figure}

\subsubsection{Web Scan}

The web assessment of CryptoKitties is shown in TABLE \ref{web_scan_Cryptokitties}. The result shows that there are 6 pages on CryptoKitties site have not set a X-Frame-Options header. This header is not included in the HTTP response to protect against ``clickjacking" attacks, which meaning attackers have chances to use a transparent ``iframe" to overlay the page and entice users to unwittingly click on the malicious options.

\begin{table}[htp]
\centering
\begin{tabular}{|p{1.5cm}|p{6cm}|}
\hline
\textbf{Medium} & X-Frame-Options Header Not Set(6) \\ \hline
\multicolumn{1}{|l|}{Method} & GET \\ \hline
\multicolumn{1}{|l|}{Parameter} & X-Frame-Options \\ \hline
\end{tabular}
\caption{Web Risks In CryptoKitties}
\label{web_scan_Cryptokitties}
\end{table}

\subsubsection{Smart Contract Audition}

Part of the assessment of CryptoKitties contracts was shown in TABLE \ref{myth_scan_Cryptokitties}.


\begin{table}[htp]
\centering
\begin{tabular}{|p{1.5cm}|p{6.5cm}|}
\hline
\multicolumn{1}{|l|}{\textbf{Low}} & \textbf{Exception State(5)} \\ \hline
\multicolumn{1}{|l|}{Function} & isPregnant(unit256)\\ \hline
\multicolumn{1}{|l|}{Function} & canBreedWith(unit256, unit256) \\ \hline
\multicolumn{1}{|l|}{Function} &  giveBirth(unit256) \\ \hline
\multicolumn{1}{|l|}{Function} &  cooldowns(unit256) \\ \hline
\end{tabular}
\caption{Smart Contract Vulnerabilities In CryptoKitties}
\label{myth_scan_Cryptokitties}
\end{table}

We detected five reachable exceptions in four categories: division by zero, out-of-bounds array access, or assert violations. Solidity uses a \textit{require()} function to check the validity of determining statement. We extract the \textit{require()} statements, which may trigger exceptions, from four functions that alerted by the Mythril.

\begin{lstlisting}[language=Solidity, numbers=none]
pragma solidity ^0.4.11;

function isPregnant(uint256 _kittyId)
    {
        require(_kittyId > 0);
        // A kitty is pregnant if and only if this field is set
    }
function canBreedWith(uint256 _matronId, uint256 _sireId)
    {
        require(_matronId > 0);
        require(_sireId > 0);
    }
function giveBirth(uint256 _matronId)
    {
        // Check that the matron is a valid cat.
        require(matron.birthTime != 0);
        // Check that the matron is pregnant, and that its time has come!
        require(_isReadyToGiveBirth(matron));
    }
function setSecondsPerBlock(uint256 secs)
    {
        require(secs < cooldowns[0]);
    }
\end{lstlisting}

After we examine the source code above, we found that in Cryptokittes,  there is actually a little risk of triggering an exception, because Variables like ``\_kittyId", ``\_matronId" or ``\_matron.birthTime" were designed to fit the requirements. For example, there are no minus options on these variables that may give them any chance smaller than zero. Thus, these "Exception State" can be defined as secure.

\subsection{0xUniverse}

0xUniverse is a blockchain game where players can build spaceships, explore the galaxy, and colonize planets. It is among the most popular blockchain game in 2019, ranked the top 3 games on Ethereum in terms of popularity.

\begin{figure}[htp]
\centering
\includegraphics[width=8.763cm]{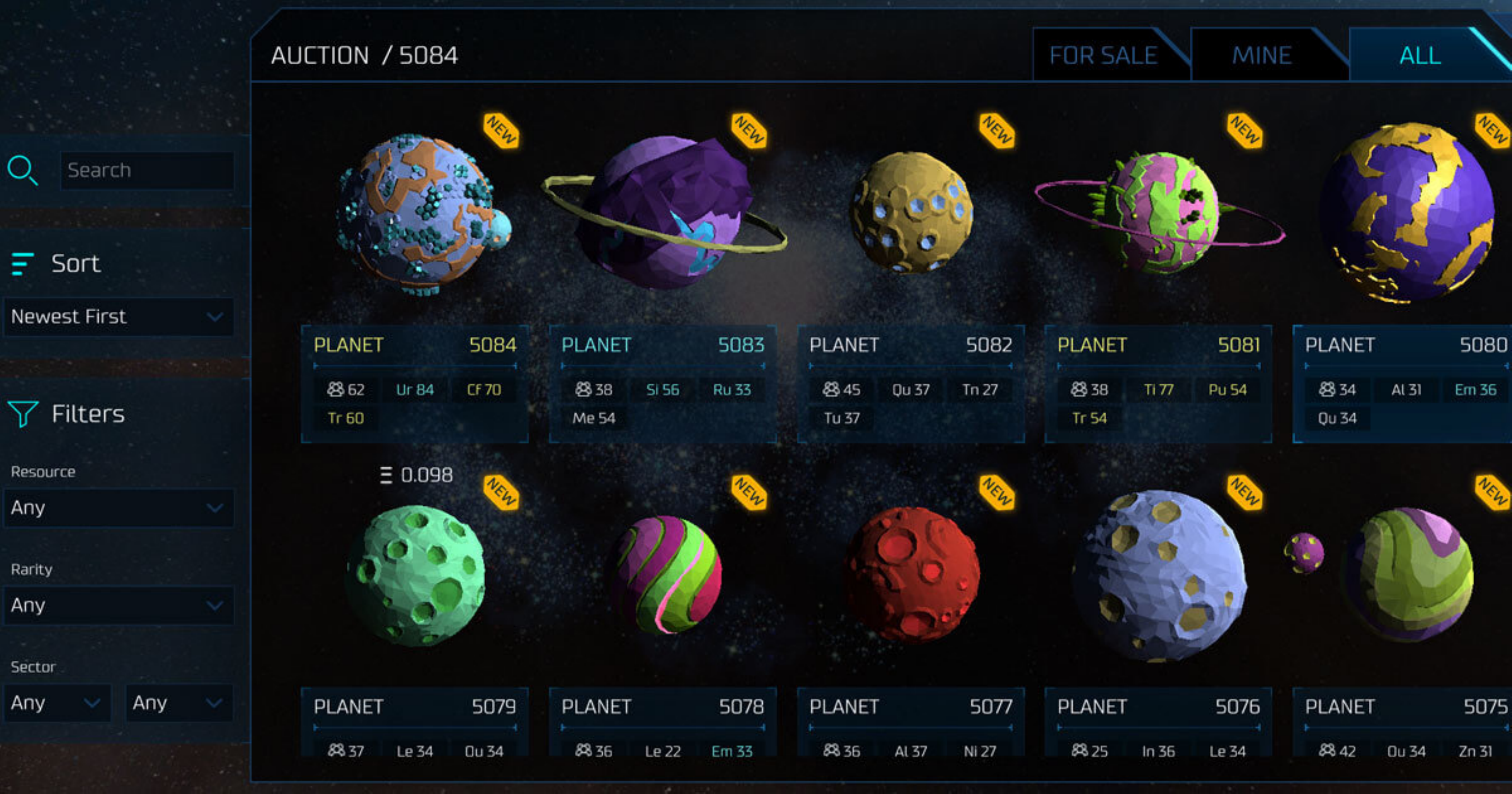}
\caption{Screenshot of 0xUniverse}\label{fig:0xuniverse}
\end{figure}

\subsubsection{Web Scan}

The web assessment of it is shown in TABLE \ref{web_scan_0xUniverse}. There are two high-risk vulnerabilities of remote OS injection, which have a potential risk of executing unauthorized operating system commands. When the web application takes in unauthorized input to OS command lines or doing improper call of external codes, there can be an OS injection. The second vulnerability, application error disclosure, which means the page contains an error message that discloses sensitive information. This sensitive information may help hackers in further attacks. However, in this case, it is a misjudgment of recognizing a included JavaScript class, tron-web\footnote{https://github.com/tronprotocol/tron-web}, as GET method's error message.

\begin{table}[htp]
\centering
\begin{tabular}{|p{1.5cm}|p{6.5cm}|}
\hline
\textbf{High} & Remote OS Command Injection(2) \\ \hline
\multicolumn{1}{|l|}{URL} & \url{https://play.0xuniverse.com/js?query=query\%22\%26sleep+15\%26\%22} \\ \hline
\multicolumn{1}{|l|}{Method} & GET \\ \hline
\multicolumn{1}{|l|}{Parameter} & query \\ \hline
\multicolumn{1}{|l|}{Attack} & query"\&sleep 15\&" \\ \hline
\multicolumn{1}{|l|}{URL} & \url{https://play.0xuniverse.com/js/blockchain/TronWeb.js?query=query\%26sleep+15\%26} \\ \hline
\multicolumn{1}{|l|}{Method} & GET \\ \hline
\multicolumn{1}{|l|}{Parameter} & query \\ \hline
\multicolumn{1}{|l|}{Attack} & query\&sleep 15\& \\ \hline
\textbf{Medium} & Application Error Disclosure(1) \\ \hline
\multicolumn{1}{|l|}{URL} & \url{https://play.0xuniverse.com/js/blockchain/TronWeb.js} \\ \hline
\multicolumn{1}{|l|}{Method} & GET \\ \hline
\multicolumn{1}{|l|}{Evidence} & Invalid parameter type \\ \hline
\textbf{Medium} & X-Frame-Options Header Not Set(1) \\ \hline
\multicolumn{1}{|l|}{Method} & GET \\ \hline
\multicolumn{1}{|l|}{Parameter} & X-Frame-Options \\ \hline
\end{tabular}
\caption{Web Risks In 0xUniverse}
\label{web_scan_0xUniverse}
\end{table}

\subsubsection{Smart Contract Audition}

As introduced in the Section~\ref{sec:architecture}, integer overflow/underflow is a common vulnerability in programming. To prevent overflow/underflow, on one hand, developers can do verification before and after the calculation, or use the SafeMath\footnote{https://github.com/OpenZeppelin/openzeppelin-solidity} library provided by OpenZeppelin. On the other hand, for functions that can trigger overflows, developers should pay more attention to authentic management.

\begin{table}[htp]
\centering
\begin{tabular}{|p{1.5cm}|p{6.5cm}|}
\hline
\multicolumn{1}{|l|}{\textbf{High}} & \textbf{Integer Overflow(1)} \\ \hline
\multicolumn{1}{|l|}{Function} & name() \\ \hline
\multicolumn{1}{|l|}{call data} & 0x06fdde03 \\ \hline
\multicolumn{1}{|l|}{call value} & 0x0 \\ \hline
\end{tabular}
\caption{Smart Contract Vulnerabilities In 0xUniverse}
\label{myth_scan_0xUniverse}
\end{table}

\subsection{Mythereum}

\begin{figure}[htp]
\centering
\includegraphics[width=8.763cm]{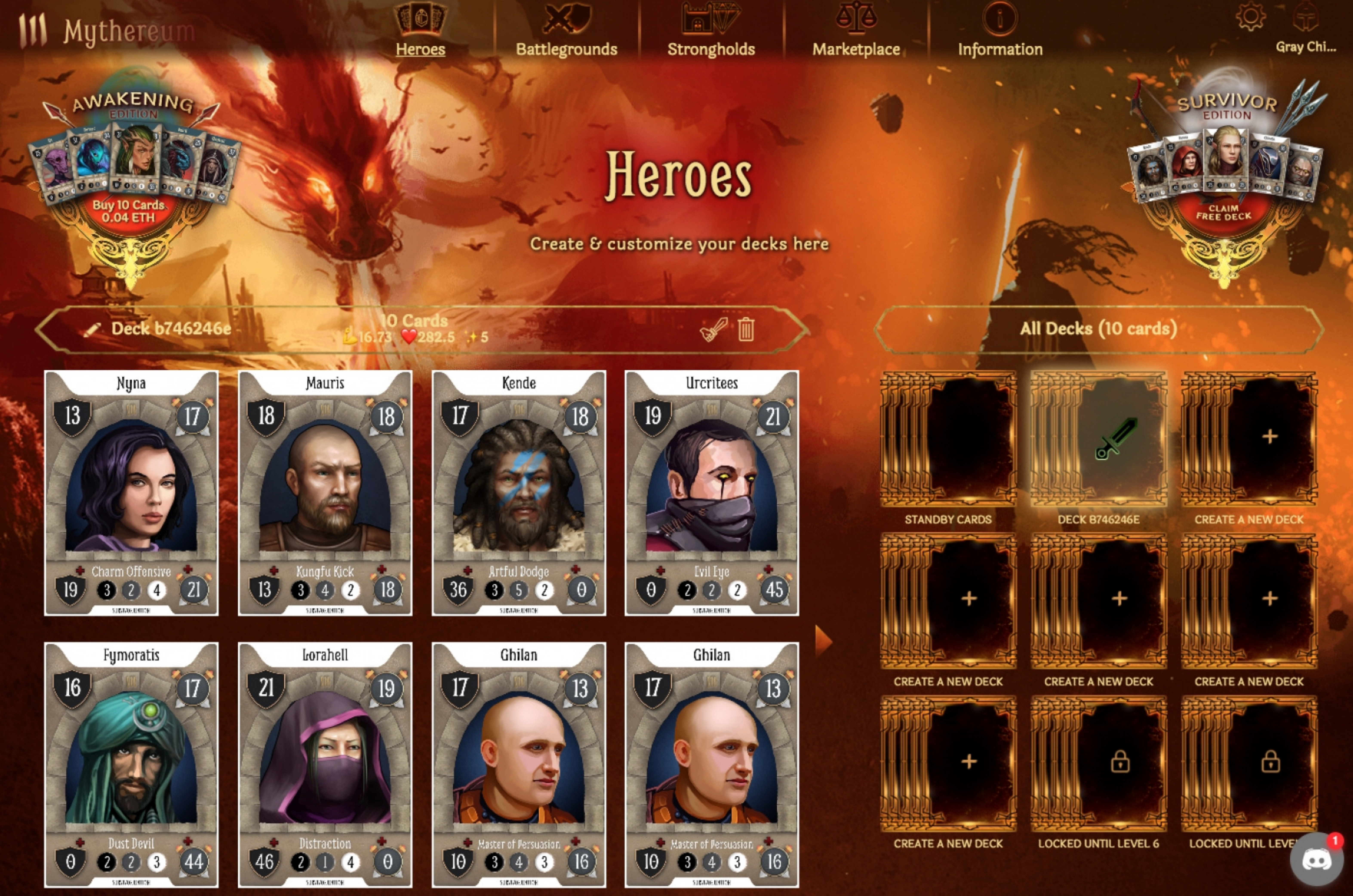}
\caption{Screenshot of Mythereum}\label{fig:Mythereum}
\end{figure}

Mythereum is a multiplayer digital trading card game built on the Ethereum blockchain where players build unique decks of collectible cards and challenge others to engage in battle. The players can launch attacks while attempting to protect their own Health and outlive every other player, earning Mythereum XP along the way.

\subsubsection{Web Scan}

The web security assessment of Mythereum is shown in TABLE \ref{web_scan_Mythereum}. We found that Mythereum is including raven.min.js from CryptoKitties. Unless the target domain name is 100\% trusted, do not use external resources on other domain names as much as possible. In our detection, Japanwar\footnote{janponwar.xyz}, another decentralized game used parlous resources is now being hijacked.

\begin{table}[htp]
\centering
\begin{tabular}{|p{1.5cm}|p{6.5cm}|}
\hline
\textbf{Medium} & Application Error Disclosure(1) \\ \hline
\multicolumn{1}{|l|}{URL} & \url{https://www.mythereum.io/main.bundle.2b361e86f68ad022f065.js} \\ \hline
\multicolumn{1}{|l|}{Method} & GET \\ \hline
\multicolumn{1}{|l|}{Parameter} & X-Frame-Options \\ \hline
\textbf{Medium} & Cross-Domain JavaScript Source File Inclusion(1) \\ \hline
\multicolumn{1}{|l|}{URL} & \url{https://cryptokitties.co} \\ \hline
\multicolumn{1}{|l|}{Method} & GET \\ \hline
\multicolumn{1}{|l|}{Parameter} & \url{https://cdn.ravenjs.com/3.20.1/raven.min.js} \\ \hline
\end{tabular}
\caption{Web Risks Mythereum}
\label{web_scan_Mythereum}
\end{table}

\subsubsection{Smart Contract Audition}

The smart contract vulnerability assessment of Mythereum is shown in TABLE \ref{myth_scan_Mythereum}.

\begin{table}[htp]
\centering
\begin{tabular}{|p{1.5cm}|p{6.4cm}|}
\hline
\multicolumn{1}{|l|}{\textbf{High}} & \textbf{Integer Underflow(2)} \\ \hline
\multicolumn{1}{|l|}{Function} & tokenURI(uint256), \_function\_0xf319428d \\ \hline
\multicolumn{1}{|l|}{call data} & 0xf319428d, 0xc87b56dd, \url{0xa22cb465000000000000000000000000010101020101010202010101010101010101010101} \\ \hline
\multicolumn{1}{|l|}{call value} & 0x0 \\ \hline
\multicolumn{1}{|l|}{\textbf{Medium}} & \textbf{Multiple Calls in a Single Transaction(6)} \\ \hline
\multicolumn{1}{|l|}{Function} & claim() \\ \hline
\end{tabular}
\caption{Smart Contract Vulnerabilities In Mythereum}
\label{myth_scan_Mythereum}
\end{table}

In the result, there was a medium vulnerability called "Multiple calls in a single transaction". When developers want to implement a payment function, they may first think of \textit{send()}. However, external calls may fail unexpectedly or intentionally. It is usually the best isolating each external call into their own transaction that can be called by the recipient. It is safer letting users withdraw funds rather than sending funds to them. Apart from that, we also found a possible underflow vulnerability shown below. Although the developer used \textit{assert()} to guarantee that minuend must be greater than the subtrahend, they didn't compare the result with the minuend. A filter should be added to ensure the result is smaller than the minuend, or simply use the SafeMath library\footnote{https://github.com/OpenZeppelin/openzeppelin-solidity/blob/master/contracts/math/SafeMath.sol}.

\begin{lstlisting}[language=Solidity, numbers=none]
  function minus(
    uint256 minuend,
    uint256 subtrahend
  ) public pure returns (uint256 difference) {
    assert(minuend >= subtrahend);
    difference = minuend - subtrahend;
  }
\end{lstlisting}

\section{Conclusion}\label{sec:conclusion}

Blockchain games benefit from the features of DApps. Compare to the infinite number of items that sold by the operators in traditional games, the items in the blockchain game are real properties, just like in the reality. Therefore, developers of blockchain games need to pay special attention to the security issue. However, according to the 2018 security report, the security situation of the DApps is still not optimistic. In this article, we introduced the background and related works on blockchain and smart contracts. Then, we discussed possible attack methods based on blockchain game architecture. After that, we illustrated overviews of the blockchain game security in terms of web application and the smart contracts. Finally, in the case studies, we demonstrated the result of security analysis on three games and described how to avoid these vulnerabilities in development.








\ifCLASSOPTIONcompsoc
\else
\fi





%




\bibliographystyle{ieeetr}
\bibliography{library}





\end{document}